\documentstyle[epsf,12pt,mycount]{article}

\newcommand{\be}{\begin{eqnarray}}     	\newcommand{\ee}{\end{eqnarray}}   
\newcommand{\rf}[1]{~(\ref{#1})}        \newcommand{\ct}[1]{$^{\cite{#1}}$}
\newcommand{\lb}[1]{\label{#1}}         \newcommand{\nn}{\nonumber}
\newcommand{\podpis}[1]{\caption{\small #1}}
\newcommand{\rav}{\,{=}\,}              \newcommand{\sh}[1]{\,{#1}\,}

\def\L{{\cal L}}   \def\U{{\cal U}}   \def\w{{\rm w}}
\newcommand{\half}{\frac{1}{2}}
\def\d{\partial}	\newcommand{\df}{\delta}
\def\^{\hat}	 	\def\_#1{\underline{#1}}
\def\fmin{$\phantom{-}$}

\newcommand{\e}[1]{\mbox{e}^{#1}}	
\newcommand{\tr}{\mathop{\mbox{Tr}}}
\newcommand{\const}{\mbox{const}}

\begin{document}

\title{\Large \bf Effective Energy Approach\\ to Collectively Quantized Systems\thanks
{This work is supported in part by funds provided by the U.S.
Department of Energy (D.O.E.) under cooperative 
research agreement \#DF-FC02-94ER40818.}}

\author{\normalsize Sergei~V.~Bashinsky\footnote{
Email address: {\tt sergei@mit.edu},\, phone:~(617)-253-5349,\, fax:~(617)-253-8674.}
}

\date{{\normalsize \it
Center for Theoretical Physics \\
Laboratory for Nuclear Science \\
and Department of Physics \\
Massachusetts Institute of Technology \\
Cambridge, Massachusetts 02139 \\
{~}\\}
{\normalsize (\,MIT-CTP-2910,~~hep-th/9910165. {~~~~} October 1999\,)}
}

\maketitle

\begin{abstract}
    We generalize effective energy variational techniques to study appropriately 
quantized solitonic field configurations. Our approach rests on collective quantization 
ideas and is specifically designed for the numerical evaluation of soliton parameters.
We employ this method to obtain the one-loop quantum corrections to the soliton mass and 
form factor.  Special attention is given to the 
regularization of the physical observables in the solitonic sector of the theory. 
The numerical implementation of the method is demonstrated for a simple one-dimensional 
scalar field example. 
\end{abstract}

\vfill
\newpage

\section{Introduction}

The dilemma between our classical visualization of a soliton as a compact
object and the plane-wave nature of its energy and momentum quantum eigenstates is as 
common  as the quantum description of any composite particle. Not surprisingly,
this problem has been approached with diversity of techniques$^{[1-6]}$
such as functional integral, canonical quantization, semi-classical and variational
procedures. 
The methods that have already been developed are able to provide reasonable
qualitative understanding of the physics behind a ``quantized soliton'' but, 
as a rule, they are analytically tractable only for the simplest models,
mostly in one space dimension.  Whenever it comes to definite predictions 
for physical quantities, many ambiguities appear$^{[7,8]}$:
consistent ultra-violet  cutoff methods, vacuum energy subtraction, and boundary conditions 
at infinity being only some of the examples.
These questions must be settled if we want the method to yield a definite
finite result.

The present paper is intended to complement the earlier works on 
soliton quantization in the following three aspects.
First, the approach that we develop naturally unifies the classical
image of a soliton as a solitary wave and the fundamental quantum-mechanical 
principles. It makes a simple connection with such more formal
views on this problem as the collective quantization methods$^{[4,5]}$
and the form-factor, ``Kerman-Klein'', approach applied
to solitons in the Goldstone and Jackiw paper$^{[2]}$.
This is demonstrated in Sections~2-4, where we review the basics of
collective quantization (Sec.~2), then introduce our variational scheme (Sec.~3),
and draw parallels to the Kerman-Klein method for calculating the soliton
form-factor (Sec.~4).

Second, during this work we have had in mind to construct a calculational method
for soliton parameters which can be practically implemented as a computer algorithm.
It then could be applied to analyze the quantum effects in
theories that are too complicated for analytical solution, but from qualitative
considerations$^{[9]}$
may possibly possess non-trivial stable solitonic states.
Much progress in this direction has already been achieved$^{[7,10]}$
in the last couple of years. However, the developed numerical methods for calculation 
of quantum corrections would not be fully satisfactory without understanding of
the special role played in the system description by the cyclic variables,
such as the soliton position in space.  We discuss the relative significance
of the modifications needed to incorporate the cyclic variables
(Secs.~3,\,6.2) and show on a simple example of a one dimensional
kink~(Secs.~6.1,\,6.2) how to calculate the one-loop quantum corrections to the physical 
parameters characterizing the soliton in its true, delocalized, ground state.

Third, we discuss theory regularization in the solitonic sector remembering that 
renormalization conditions are conventionally imposed in the perturbative sector of 
the theory~(Sec.~5).  In dealing with this subtle issue, we prefer to apply 
physically realizable 
regularization methods rather than formal manipulations with divergent expressions.
Our final formulas involve only finite quantities and convergent integrals,
suitable for the numerical computations of Section~6.
A summary and conclusions are given at the end of the paper~(Sec.~7).

\section{Collective quantization}

We assume that the theory has an absolutely stable solitonic state.
That is, the decay of this soliton into plane-wave excitations over the usual vacuum 
is forbidden by conservation of some charge $Q$, which could be
a topological charge or a fermion number. Different values of the charge $Q$ split 
the overall Hilbert space of the theory into separate non-mixing
sectors with their own ground states, which would be the true vacuum for the 
``perturbative'' sector and the soliton at rest for the ``solitonic'' sector.
Due to translational invariance of the theory, the total momentum operator~$\^ {\bf P}$ 
commutes with the Hamiltonian,
\be
[\^ {\bf P},\^ H]=0~,
\nn\ee
and the lowest energy state in the solitonic sector should also be an eigenstate 
of the momentum operator, corresponding to the total momentum zero.
By the Heisenberg uncertainty principle, the position of the soliton in space
is completely undefined. 
The techniques to handle this situation are known$^{[1-6]}$.

To be specific, we illustrate the formalism on the simplest example of a real scalar field
$\phi$ in one space dimension with the Lagrangian density 
\be
\L(x,t)=\half\left(\frac{\d \phi}{\d t}\right)^2{-}~\half\left(\frac{\d \phi}{\d x}\right)^2
{-}~g^{-2}\,\U(g\phi)~.
\lb{Lagr}
\ee
This theory may support a stable topological soliton if the global minimum of the
potential $\U$ is not unique:
\be
\exists~\sigma_1,\,\sigma_2:\quad \U(\sigma_1)=\U(\sigma_2)=\min\,\U~,\qquad
\sigma_1\not=\sigma_2~
\nn\ee
(we assume that the minima~$\sigma_1$ and~$\sigma_2$ are connected by a discrete
symmetry of the Lagrangian\rf{Lagr}, {\it e.g.\/} $\phi\sh{\to} {-}\phi$\,.) 
Let $\sigma(x)$ be an arbitrary function such that
\be
\lim_{x\to-\infty}\sigma(x)=\sigma_1~~,\quad
\lim_{x\to+\infty}\sigma(x)=\sigma_2~.
\lb{sigmabc}
\ee
Let also $\{\eta_k(x)\}$ be a set of orthonormal functions,
\be
\int dx~\eta_k(x)\,\eta_l(x)=\delta_{kl}~,
\nn\ee
that are orthogonal to $\sigma'(x)$,
\be
\int dx~\sigma'(x)\,\eta_k(x)=0~,
\nn\ee
and, together with $\sigma'(x)$, form a complete set.

Following the Christ and Lee method$^{[4]}$, we describe the solitonic sector of the 
theory
trading the field $\^\phi(x)$ for an equivalent set of commuting dynamical variables 
$\{\^ X,\^ q_1,\^ q_2,...\}$ as
\be
\^\phi(x)=g^{-1}\sigma(x-\^ X)+\sum_{k=1}^{\infty}\^ q_k\eta_k(x-\^ X)~.
\lb{phidec}
\ee
Given a field configuration $\phi(x)$, the corresponding variables $\{X,q_1,q_2,...\}$ 
may be determined from the equations
\be
\int dx~\sigma'(x-X)\phi(x)=0 ~,~~~~~~~~~~~~~~~~~~~~~~~~~ \lb{x0}\\
q_k=\int dx~\eta_k(x-X)\left[\phi(x)-g^{-1}\sigma(x-X)\right]~.\lb{qk}
\ee
Let $\{\^ P,\^\pi_1,\^\pi_2,...\}$ be the operators of the canonical momenta conjugate to 
$\{\^ X,\^ q_1,\^ q_2,...\}$, so that
\be
[\^ P,\^ X]=-i~,~~~[\^\pi_k,\^ q_l]=-i\delta_{kl}~
\lb{com}
\nn\ee
with all other possible commutators being zero.
The original Hamiltonian,
\be
H[\^\pi(x),\^\phi(x)]=\half\int dx\,\left(\^\pi(x)\right)^2+V[\^\phi]~
\lb{H}
\ee
with
\be
V[\phi]\equiv\int dx\left[\half\left(\phi'(x)\right)^2+
g^{-2}\,\U(g\phi(x))\right]~,
\lb{V}
\ee
when being re-expressed in terms of the new variables and their conjugate 
momenta, reads$^{[4,11]}$:
\be
H(\^ P,\^\pi,\^ q)&=&\half\sum_{k=1}^{\infty}\^\pi_k^2+V(\^ q)+
g^2 h(\^ P,\^\pi,\^ q)~. 
\lb{Hq}
\ee
In this expression $h(\^ P,\^\pi,\^ q)$ is an $O(g^2)$ correction to
the kinetic term that is quadratic in the canonical momenta $\^P$ or~$\^\pi$,
but has somewhat intricate $q$-dependence$^{[4]}$,
and $V(q)$ is simply the potential term\rf{V} written in terms of the new variables.

The Hamiltonian\rf{Hq} does not depend on the ``collective coordinate''~$\^ X$
to any order in the coupling $g$ reflecting the system invariance to translation 
of the field $\phi(x)$ in space as a whole.
Its conjugate momentum $\^ P$ is, therefore, a conserved quantity:
\be 
[\^ P,\^ H]=0~. 
\nn\ee
It is possible to show$^{[4,11]}$
that the operator $\^ P$ indeed represents the 
{\it total momentum} of the system\footnote{
Let us provide a simple illustration of this statement observing that the commutation 
relations
\be
[\^ P,\^ q_k] = 0~~~\mbox{and}~~~
[\^ P,f(\^ X)] = -if'(\^ X)~~~\forall~\mbox{function~}f(x)
\nn\ee
being applied to eq.\,(\ref{phidec}) produce immediately the formula for the total momentum 
commutator:
\be
[\^ P,\^\phi(x)]=i\frac{\d}{\d x}\^\phi(x)~.
\nn\ee
}.
The essence of the collective quantization is to describe the original theory
restricted to the subspace of the quantum states $|\Omega_P\rangle$ that carry 
a {\it definite} total momentum $P$, for example
\be
\^ P\,|\Omega_0\rangle=0~.
\nn\ee
The zero value of the momentum can always be achieved by a proper
choice of the Lorentz frame\footnote{
Quantization in moving frames may also be considered$^{[12]}$.
}. 
In the $P\rav 0$ sector the full Hamiltonian\rf{Hq} reduces to
\be
H'(\^\pi,\^ q)&=&\half\sum_{k=1}^{\infty}\^\pi_k^2+V(\^ q)+O(g^2)~, 
\lb{Hred}
\ee
and the remaining dynamical variables may be handled by conventional perturbative 
or semi-classical methods.

We would like to conclude this section by the following remark:
In the original Christ and Lee method the profile function
$g^{-1}\sigma(x)$ in eq.\rf{phidec} is required to satisfy the classical field equations of 
motion. In the present discussion, $\sigma(x)$ is an {\it arbitrary} function with
the proper asymptotic behavior, eq.\rf{sigmabc}, yielding a non-degenerate 
transformation $\phi(x)\to\{X,q_1,q_2,...\}$\,.  We exploit this freedom in the 
choice of $\sigma$ soon.

\section{Reduced effective energy}

For the solitonic sector of the theory and a given function $\sigma(x)$ in
the canonical transformation\rf{phidec}, we define
\be
\e{iW'[J_1(t),J_2(t),...]}&\equiv& \frac1{\left(\int [dq]~\e{iS'[q]}\right)}
\int [dq]~\e{iS'[q]+i\sum_{k=1}^{\infty}\int dt\,J_k(t)q_k(t)}~, \lb{Wprime}\\
\Gamma'[q_1(t),q_2(t),...]&\equiv& \min_{\{J_k(t)\}}\left(-W'[J]+
                     \sum_{k=1}^{\infty}\int dt~J_k(t)q_k(t)\right), \lb{Gprime}
\ee
where $S'[q]\,{\equiv}\,S'[q_1(t),q_2(t),...]$ in eq.\rf{Wprime} is the classical 
action of our system in the $P\,{=}\,0$ sector corresponding to the ``reduced'' 
Hamiltonian\rf{Hred}, and the functional measure $[dq]\rav\prod_{k=1}^{\infty}[dq_k]$ 
absorbs all the factors that may have appeared in the functional integral when the 
canonical momenta $\pi_k$ 
were integrated out.  Eqs.\rf{Wprime} and\rf{Gprime} are reminiscent of the standard 
definitions of the generating functional $W[J(x)]$ and the effective action 
$\Gamma[\phi_0(x)]$, when the source terms would have the form $\int d^2x~J(x)\phi(x)$.
Here, we couple the sources $J_k$ to the complicated {\it non-linear} functionals 
$q_k[\phi(x)]$. We refer the quantity $\Gamma'$ as the ``reduced effective action''. 

Let us consider the special case when the arguments of $\Gamma'$ are constant functions
$q_1,q_2,...$\,, and let the time integrals in eqs.~(\ref{Wprime}--\ref{Gprime}) be taken 
over a large but finite interval $0\sh{\le}t\sh{\le}T$. Then
\be
\Gamma'[q_1(t),q_2(t),...]=E'_{\rm eff}(q_1,q_2,...)~T~,
\nn\ee
where the {\it reduced effective energy} $E_{\rm eff}$ allows a nice 
physical interpretation$^{[13]}$.
Namely, it is the minimum of the Hamiltonian 
expectation value among the quantum states $|\Omega'\rangle$ for which 
the quantum expectation values of the operators $\{\^ q_k\}$ are given by 
$E'_{\rm eff}$~arguments,~$\{q_k\}$:
\be
E'_{\rm eff}(q_1,q_2,...)=\min_{{|\Omega'\rangle\mbox{ \tiny such that}\atop
\langle\Omega'|\^q_k|\Omega'\rangle=q_k}}
\frac{\langle\Omega'|\^H'|\Omega'\rangle}{\langle\Omega'|\Omega'\rangle}~.
\lb{Epi}
\ee

We primed the ket-vectors $|\Omega'\rangle$ in order to emphasize that these states
belong to the Hilbert space associated with the reduced Hamiltonian $\^ H'$.
It is different from the space of the states, $\{|\Omega\rangle\}$, of the original theory
described by $\^ H$ because the latter carry an additional quantum number -- the total 
momentum $P$.  However, by construction
\be
H'(\^\pi,\^q)=\left. H(\^P,\^\pi,\^q) \right|_{\^P=0}~,
\nn\ee
and one can rewrite eq.\rf{Epi} in terms of the objects of the original theory:
\be
E'_{\rm eff}(q_1,q_2,...)=\min_{{|\Omega\rangle\mbox{ \tiny such that}\atop
\langle\Omega|\^q_k|\Omega\rangle=q_k,\,\^P|\Omega\rangle=0}}
\frac{\langle\Omega|\^H|\Omega\rangle}{\langle\Omega|\Omega\rangle}~.
\lb{Epi1}
\nn\ee
Provided that we are able to compute the reduced effective energy $E'_{\rm eff}$,
the mass of the soliton in its ground state, $M$, may be determined as 
\be
M=\min_{q_1,q_2,...}E'_{\rm eff}(q_1,q_2,...)=
\min_{{|\Omega\rangle\mbox{ \tiny such that}\atop\^P|\Omega\rangle=0}}
\frac{\langle\Omega|\^H|\Omega\rangle}{\langle\Omega|\Omega\rangle}~.
\lb{min_E}
\ee

Now we proceed to the actual computation of $E'_{\rm eff}$. Our first crucial step is
to trade the discrete infinite set of variables $\{q_1,q_2,...\}$ for a more physically
intuitive continuous ``shape'' function.  As eq.\rf{min_E} suggests, the mass of the
soliton, in principle, could be calculated by searching for the minimum of $E'_{\rm eff}$ 
over the range of parameters $\{q_1,q_2,...\}$ that are defined by means of the 
decomposition\rf{phidec} with an arbitrarily chosen but fixed function $\sigma(x)$.
However, we find it more straightforward to always calculate $E'_{\rm eff}$ at
zero expectation values of all $\^ q_k$ and vary the function $\sigma(x)$ itself
looking for the soliton mass as
\be
M=\min_{\sigma(x)}E'_{\rm eff}[\sigma(x)]~.
\lb{min_sigma}
\ee 
In this equation, $E'_{\rm eff}[\sigma(x)]$ is $E'_{\rm eff}(0,0,...)$, by our previous 
notations,
where the coordinates $\{q_k\}$ are defined using the given ``shape'' function~$\sigma(x)$.
As shown in Section~4, the configuration~$\sigma(x)$ that minimizes the
right hand side of eq.\rf{min_sigma} is closely connected to the soliton form factor.

The semi-classical expansion for the reduced effective energy 
\be
E'_{\rm eff}[\sigma(x)] = E_{\rm cl}[\sigma(x)] + E'_{\rm Cas}[\sigma(x)] + O(g)
\lb{Esum}
\ee
starts from an $O(g^{-2})$ term -- the classical energy associated with the field 
configuration $\phi(x)\rav g^{-1}\sigma(x)$:
\be
E_{\rm cl}[\sigma] = V[g^{-1}\sigma] = \frac1{g^2}\int dx 
\left[\frac12\left(\sigma'(x)\right)^2 + \U\left(\sigma(x)\right)\right]~.
\lb{Eclass}
\ee
The leading quantum correction, $\frac{i}{T}\ln\det\left.\left(
\frac{\df^2 S'[\sigma(x),q]}{\df q_k \df q_l}\right)\right|_{q=0}\rav\, O(g^0)$, 
can be written$^{[14]}$
(up to $O(g)$ terms neglected in the one-loop approximation)
as the ``Casimir energy''
\be
E'_{\rm Cas}[\sigma(x)] = \frac{1}{2}\left(\sum_{k=1}^{\infty} 
\omega'_k[\sigma(x)]\right)_{\rm regularized}~,
\lb{ECas}
\ee
where $\{\omega'_k\}_{k=1,2,...}$ by definition are the frequencies of small 
oscillations in the reduced system ``stabilized'' at $q\rav 0$ by an
``external source''. That is, $\{\omega'_k\}$ are the normal frequencies of
the Hamiltonian 
\be
\tilde H'(\pi,q)&=&H'(\pi,q)-
\sum_{k=1}^{\infty}\left(\left.\frac{\d H'}{\d q_k}\right|_{q=0}\right)q_k=\\
&=&\half\sum_{k=1}^{\infty}\pi_k^2+V(q)+O(g^2)-
\sum_{k=1}^{\infty}\left(\left.\frac{\d V}{\d q_k}\right|_{q=0}\right)q_k~.
\lb{Hauxd}
\ee

We claim that, with the accuracy including at least $O(g)$, the frequencies 
$\{\omega'_k\}$ are given by (the square roots of) the {\it non-zero} 
eigen-values in the following Sturm-Liouville problem:
\be
\left(\omega'_{n}[\sigma]\right)^2\eta_n(x)&=
&\left[-\,\frac{d^2}{d x^2}+\U''(\sigma)\right]{\eta}_n(x)+
   \int dy\,\left\{\frac{j_{\sigma}'(x)\sigma'(y)+
 \sigma'(x)j_{\sigma}'(y)}{\left[\int dz\,\left(\sigma'(z)\right)^2\right]}\right.-\nonumber\\
&~&\qquad\qquad\qquad -\left.\frac{\left[\int dz~j_{\sigma}'(z)\sigma'(z)\right]}
      {\left[\int dz\,\left(\sigma'(z)\right)^2\right]^2}~\sigma'(x)\sigma'(y)
\right\}\eta_n(y)~ \lb{nonloc2}
\ee
where
\be
j_{\sigma}(x)\equiv - g\left.\frac{\df H}{\df\phi(x)}\right|_{\phi=g^{-1}\sigma}=
          \sigma''(x)-\U'\left(\sigma(x)\right)~.
\lb{source}
\ee
One of the several methods to prove this statement consists in modifying the kinetic 
term in the ``stabilized'' reduced Hamiltonian\rf{Hauxd} by addition of the $P$-dependent 
piece $g^2h(P,\pi,q)$ that appears in
eq.\rf{Hq}. Then we transform the canonical variables back to the continuous 
fields $\phi(x)$ and~$\pi(x)$, when eq.\rf{qk} comes helpful. As the result of this 
procedure,
\be
\tilde H'(\pi,q) &\rightarrow& \frac12\sum_{k=1}^{\infty}\pi_k^2+V(q)+g^2h(P,\pi,q)-
\sum_{k=1}^{\infty}\left(\left.\frac{\d V}{\d q_k}\right|_{q=0}\right)q_k= 
\lb{modification}\\
&=&H[\pi(x),\phi(x)] + \frac1g\int dx~j_{\sigma}(x)\,\phi(x+X[\phi])\,\equiv\,
\tilde H[\pi(x),\phi(x)]~,
\lb{Haux}
\ee   
where $H[\pi(x),\phi(x)]$ is the Hamiltonian of the original theory\rf{H} and
the second term in eq.\rf{Haux} is the transform of the stabilizing
term from the previous expression. 
The non-linear, {non-local} functional $X[\phi(x)]$ in eq.\rf{Haux} is explicitly 
defined by eq.\rf{x0} and the source $j_{\sigma}(x)$ is precisely the one that appears in 
eq.\rf{source}.  It is not hard to argue that 
the modification\rf{modification} does not affect the spectrum of small oscillations
about the classical equilibrium point of this system more than 
$O(g^2)$, although it implies an additional degree of freedom associated with
the momentum~$P$ and the oscillation spectrum of $\tilde H[\pi(x),\phi(x)]$ should 
acquire an additional eigen-mode.  However, the new canonical variable~,``$X$'', 
does not explicitly
enter~$\tilde H$ as it was not present in the previous expression\rf{modification},
leaving the Hamiltonian\rf{Haux} still being invariant under 
$\phi(x)\sh{\to}\phi(x\sh{+}a)$\,, and
the eigen-frequency of the new, translational, mode being identically {\it zero\/} for any 
configuration~$\sigma(x)$.  Finally, the oscillation spectrum of $\tilde H$ is determined 
by the eigen-value problem
\be
\left(\omega'_{n}\right)^2\eta_n(x)&=&\int dy\,\left(\left.\frac{\df^2\tilde H}{\df\phi(x)
  \df\phi(y)}\,\right|_{\phi=g^{-1}\sigma}\right)\eta_n(y)~
\nn\ee
that after some algebra leads to eq.\rf{nonloc2} above.  

To summarize, the one-loop contribution to the effective energy in the $P\rav 0$ sector
is given by the sum\footnote{Since $\omega'_0\sh{\equiv}0$, it does not matter whether
one includes it in the sum or leaves out. 
}
of $\frac{1}{2}\,\omega'_n[\sigma]$ that are 
determined by the Schroedinger-type equation~(\ref{nonloc2}) with a non-local 
but {\it separable\/} 
potential. This sum is terribly divergent and should be regularized by subtracting
the Casimir energy of the trivial vacuum and proper counterterms specified by 
renormalization conditions in the perturbative sector.
We perform this regularization in Section~5.
In the absence of the non-local terms in eq.\rf{nonloc2}, the sum of the corresponding
eigen-modes, including $\omega_0$, would yield the one-loop contribution to the 
conventional effective energy associated with the spatially ``nailed down'' field 
configuration $\phi(x)\rav g^{-1}\sigma(x)$.  The non-local terms vanish for 
$\sigma(x)\rav \sigma_{\rm cl}(x)\sh{\equiv} g\,\phi_{\rm cl}(x)$
that solves the classical Euler-Lagrange equation, 
\be
\frac{d^2\sigma_{\rm cl}}{d x^2}-\U'(\sigma_{\rm cl})= 0~,
\lb{EL}
\ee
because then $j_{\sigma_{\rm cl}}(x)\sh{\equiv}0$ by its definition\rf{source}.
These terms guarantee that eq.\rf{nonloc2} allows a normalizable zero-frequency solution 
$\eta_0(x)\sh{\propto}\sigma'(x)$ even when $\sigma(x)$ differs from the classical
configuration. However, the effect of the non-local terms on the frequencies
with $n\sh{\not=}0$ is only quadratic in $\delta\sigma \equiv \sigma\sh{-}
\sigma_{\rm cl}$\,. Indeed, calculating  $\delta\omega'\sh{\equiv} \omega'[\sigma]\sh{-}
\omega'[\sigma_{\rm cl}]$ by the usual perturbation theory, we see these terms do not 
contribute in the first order because $j_{\sigma_{\rm cl}}\sh{\equiv} 0$ and 
$\int dz\,\sigma'\eta_n\rav 0$ when $n\sh{\not=}0$.

Minimizing the reduced effective energy\rf{Esum} with respect
to all possible configurations $\sigma(x)$ one finds the mass of the soliton, $M$\,. 
A consistent approximation scheme for this procedure can be implemented as a power 
series in the coupling $g$ that starts from 
\be
M_{\rm cl} = \min E_{\rm cl} = E_{\rm cl}[\sigma_{\rm cl}] = O(g^{-2})~.
\nn\ee 
The next, Casimir, term modifies the mass by 
\be
\delta M = E'_{\rm Cas}[\sigma_{\rm cl}] = O(g^0)~.
\lb{dM}
\ee 
Since, in general,
\be
\left.\frac{\delta E'_{\rm Cas}[\sigma]}{\delta\sigma}\right|_{\sigma_{\rm cl}} =   
\frac{1}{2}\left(\sum_{k=1}^{\infty}\left.\frac{\delta\omega'_k[\sigma]}
{\delta\sigma}\right|_{\sigma_{\rm cl}}\right)_{\rm regularized}\not= 0~,
\lb{Cas_lin}
\ee 
the Casimir term in $E'_{\rm eff}$ also changes the minimizing configuration, or the
soliton form factor, by $O(g^2)$.   However, according to the previous paragraph,
the non-local potential in eq.\rf{nonloc2} does not affect the quantity\rf{dM}
or the sum in eq.\rf{Cas_lin}.
We {\it conclude} that for the purpose of calculating the first quantum correction 
to the classical values of the soliton mass and form factor it is sufficient to 
sum up the oscillation frequencies $\omega_n$ about a spatially fixed configuration
$\phi(x)$, omitting the non-local terms in eq.\rf{nonloc2}, but {\it throw away}
the lowest eigen-frequency, $\omega_0$\,.

Nevertheless, eqs.~(\ref{ECas},\,\ref{nonloc2}) allow one to compute the well defined 
one loop effective energy in the $P\rav 0$ sector for the configurations not 
necessarily close to the solution of the classical equations of motion. This can be helpful
in the search for non-perturbative radiatively generated solitons$^{[15]}$,
at least 
on the qualitative level.  Therefore, in our calculational example in Section~6 we exploit 
the full form of eq.\rf{nonloc2} including all the non-local terms. 

\section{Shape Function and the Soliton Form Factor}

In this short Section we demonstrate that, to our one-loop accuracy,
the function $\sigma(x)$ minimizing the functional 
$E'_{\rm eff}[\sigma]$ has a physical interpretation of the soliton form-factor.
Indeed, let us evaluate the matrix element of the operator $\^\phi(x)$ between the stable 
solitonic states with the mass $M$ that differ by their total momenta:
\be
\langle P'|\^\phi(x)|P\rangle = 
\e{-i(P'-P)x}\langle P'|\^\phi(0)|P\rangle~.
\lb{bffd}
\ee

First, we present the state $|P\rangle$ as the soliton at rest, $|0\rangle$\footnote{
The state $|0\rangle$ should {\it not} be confused with the true, perturbative, vacuum.
}, boosted to 
the momentum~$P$ by the Lorentz boost operator $\^\Lambda_P$ and recall
the scalar nature of the field~$\^\phi(x)$. We have
\be
\langle P'|\^\phi(0)|P\rangle = \langle P'|\^\phi(0)\^\Lambda_P|0\rangle =
\langle P'|\^\Lambda_P\^\phi(0)|0\rangle = \langle Q|\^\phi(0)|0\rangle = f(Q)~, 
\nn\ee
where
\be
|Q\rangle = \^\Lambda^{\dagger}_P |P'\rangle = \^\Lambda_{-P} |P'\rangle
\nn\ee
describes a soliton carrying the momentum
\be
Q=\frac{P'E-PE'}{M}=(P'-P)-\frac{PP'}{2M^2}(P'-P)+...= P'-P +O(g^4)~.
\nn\ee
For the very last estimate, we remember that $M\rav O(g^{-2})$ and that we are interested 
in the momenta $P$ and $P'$ of the order of the inverse soliton size that is $O(g^0)$\,.

Second, we substitute $\^\phi(x)$ by the decomposition\rf{phidec} and Fourier transform 
the functions $\sigma(x)$ and $\eta(x)$: 
\be
f(Q) &\equiv& \langle Q|\^\phi(0)|0\rangle = 
g^{-1} \langle Q| \sigma(-\^ X)+g\,\^q_k\eta_k(-\^ X)|0\rangle = \nonumber\\
&=& g^{-1} \int\frac{dQ'}{2\pi}\langle Q| \e{iQ'\^ X}
   \left[\tilde\sigma(Q')+g\,\^q_k\tilde\eta_k(Q')\right] |0\rangle~.
\lb{effd}
\ee

Third, let us notice that the state 
$\e{-iQ'\^ X}|Q\rangle$,  being a momentum eigenstate because
\be
\^ P \left(\e{-iQ'\^ X}|Q\rangle\right) = \e{-iQ'\^ X}(\^ P -Q')|Q\rangle = 
(Q-Q')\left(\e{-iQ'\^ X}|Q\rangle\right)~,
\nn\ee  
is also almost an eigenstate of the Hamiltonian\rf{Hq}\,:
\be
\^ H \left(\e{-iQ'\^ X}|Q\rangle\right) = \e{-iQ'\^ X} \^ H |Q\rangle + 
[\^ H,\e{-iQ'\^ X}] |Q\rangle = M \left( \e{-iQ'\^ X}|Q\rangle \right)
 + \^O(g^2)|Q\rangle~. 
\nn\ee
From these two observations we conclude that 
\be
\e{-iQ'\^ X}|Q\rangle = \left(1+O(g^4)\right)|Q-Q'\rangle + 
                        O(g^2)|(Q-Q')^*\rangle~,
\lb{star}
\ee
where $|(Q-Q')^*\rangle$ describes an excited soliton or a soliton-meson
scattering state with the total momentum $Q\sh{-}Q'$\,.
The leading term in eq.\rf{star} is, of course, expected to be the same soliton 
boosted to the momentum $Q\sh{-}Q'$
for a heavy, non-relativistic particle described by the 
operators $\^X$ and~$\^ P$.  The higher order corrections should be anticipated as well
because $\e{iP\^ X}$ is not the complete boost operator in the field theory. 

Substituting the result\rf{star} in eq.\rf{effd} and applying the normalization 
conditions\footnote{
   The normalization in eq.\rf{norm} is taken to agree with 
the paper~\cite{GJ}. The second equation holds automatically for the forward 
matrix element, $\langle 0|\^ q_k|0\rangle$, by our choice of the configuration $\sigma(x)$
that keeps the expectation values of $q$'s at zero
(see the text around eq.\rf{min_sigma}).  For the off-forward matrix~elements,
\be
\langle Q-Q'|\^ q_k|0\rangle = \frac{1}{Q-Q'}\,\langle Q-Q'|[\^ P,\^ q_k]|0\rangle = 0~,
\qquad
Q-Q'\not= 0~. 
\nn\ee
}
\be
&&\langle Q-Q'|0\rangle = 2\pi\,\delta(Q-Q')~, \lb{norm}\\
&&\langle Q-Q'|\^ q_k|0\rangle = 0~,           \lb{q_exp}
\ee
one arrives at the formula
\be
f(Q) = g^{-1} \left\{\tilde\sigma(Q) + O(g^3)\right\}~.  
\lb{fQ}
\ee

Summarizing eqs.~(\ref{bffd}--\ref{fQ}),
\be
g\,\langle P'|\^\phi(x)|P\rangle &=& 
\e{-i(P'-P)x} \left( \tilde\sigma(P'-P) + O(g^3) \right) = \\ 
&=&\int dy~\e{-i(P'-P)y}\sigma(x-y) + O(g^3)~~,
\lb{formfr}
\ee
which is the ansatz~(ii) of Goldstone and Jackiw$^{[2]}$
corrected for the one-loop
quantum effects.  In the perturbative expansion about the classical configuration,
\be
\sigma(x)=\sigma_{\rm cl}(x) + \delta\sigma(x)~,
\lb{chert}
\ee
we obtain the following linear equation for the leading order correction $\delta\sigma$ 
that minimizes the sum $E_{\rm cl}\sh{+}E'_{\rm Cas}$: 
\be
\left[-\frac{d^2}{dx^2}+\U''(\sigma_{\rm cl}(x))\right]\delta\sigma(x) =
-g^2\sum_{k=1}^{\infty}\frac{\U'''\left(\sigma_{\rm cl}(x)\right)\eta^2_k(x)}
     {4\,\omega'_k[\sigma_{\rm cl}]}\,+\,O(g^3)~.
\lb{d_sigma}
\ee
This is the formula for the one-loop correction to the soliton form-factor derived
within the Kerman-Klein method$^{[2]}$.
The non-local terms in eq.\rf{nonloc2}
take care of the (would be infinite) zero-mode term on the right hand side of 
eq.\rf{d_sigma} but give no other contribution in this order of the
perturbative expansion\rf{chert}, as it was already stated at the end of Section~3. 

\section{Regularization}
 
Now we return to the main line of our discussion and consider the reduced Casimir energy, 
currently written in the form\rf{ECas}. The aim of this section is to convert this formal 
expression into an unambiguous and finite quantity calculable numerically with our 
computer. To avoid the many possible pitfalls in this step, we attempt to be very
explicit in details.
First of all, in order to regulate the $\omega'_k[\sigma]$ sum in eq.\rf{ECas},
one should subtract the Casimir energy of the true, perturbative, vacuum.   For the 
scalar field $\phi(x)$ living in a box of length $L$ with fixed $\phi$ values at the  
boundaries, the normal oscillation modes (``meson'' excitations) about the topologically 
trivial minimum energy configuration $\phi(x)\,{\equiv}\,g^{-1}\sigma_1$ (or $\sigma_2$) 
have the frequency  
\be
\omega^{(0)}_n=\sqrt{m^2+k^2_{n}}\equiv
\sqrt{\U''(\sigma_{1})+\left(\frac{\pi(n+1)}{L}\right)^2}\quad,\qquad n=0,1,\dots~.
\lb{vac_mod}
\ee
Each of the sums $\sum\omega'_k[\sigma]$ and $\sum\omega^{(0)}_n$ diverges badly  
and one should specify an ultra-violet cutoff prescription. It must be imposed
consistently in both the trivial and the solitonic sectors\footnote{
It would be dangerous, and actually wrong in one space dimension,
to cut off the oscillation modes at some fixed energy which is the same in both sectors 
because the number of modes below the cutoff in these sectors generally differs.
}.
This is not straightforward to achieve because of many incongruities between both the 
system description and its spectrum in the two sectors. 

Let us keep in mind that the ultra-violet divergences are unambiguously 
regularized by replacement of the continuous field $\phi(x)$ with a large but finite 
number of degrees of freedom having the same low-energy behavior, as provided, 
for example, by a lattice version of the theory. 
We imagine a continuous and physically 
realizable process that adiabatically transfers the regularized (quantum!) system from the 
vacuum state to a solitonic one by some external influence. 
As a specific example, we consider the $\phi^4$ model\footnote{
The coefficients in eq.\rf{Uphi4} are chosen so that $\sigma_{1,2}\rav{\mp}1$ and 
$\U''(\sigma_{1,2})\sh{\equiv}m^2\rav1$.
}
\be
\U(\sigma)=\frac18\left(\sigma^2-1\right)^2~,
\lb{Uphi4}
\ee
where we confine the scalar field $\phi(x)$ to a finite box $0\sh{\le}x\sh{\le}L$\,,
$L\gg 1$ and impose the boundary conditions
\begin{figure}[t]
\centering \leavevmode
\epsfbox{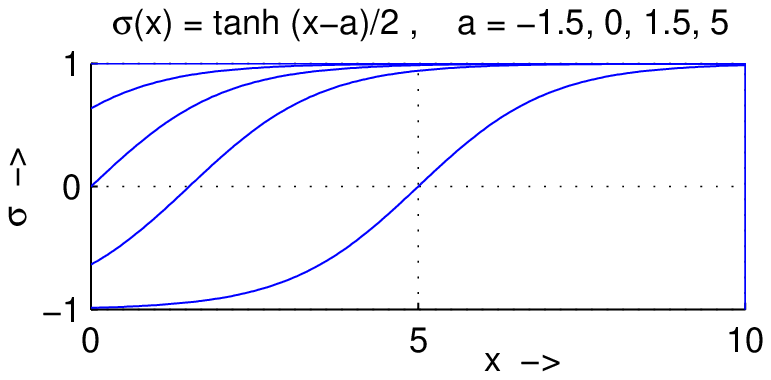}
\podpis{Classical solutions of the Euler-Lagrange equation for a scalar field in the
box $0\sh{\le} x \sh{\le}10$ and the $\phi^4$ potential, eq.\rf{Uphi4}.
The field satisfies the boundary conditions $\sigma(0)\rav s \sh{\in}[-1,1]$
and $\sigma(10)\sh{\simeq} 1$. The connection between the parameter $s$ and the kink
position, $a$, is given by eq.\rf{as}.
} 
\label{fig1}
\end{figure}
\be
\left\{ \begin{array}{l} 
\phi(0)= g^{-1}s~~,\quad -1\leq s \leq 1~;\\
\phi(L)= g^{-1}~
\end{array} \right.
\lb{bc}
\ee
that smoothly interpolate between the trivial and non-trivial topologies as the
parameter $s$ varies from $1$ to $-1$. We would like to examine the behavior of the classical
oscillation frequencies and the quantum Casimir energy in the ground state of this
system as a function of $s$.
The classical time-independent field equation\rf{EL}
with the potential\rf{Uphi4} admits the solutions
\be
\sigma(x)\equiv g\phi(x) = \tanh\frac{x-a}{2}~, \qquad \forall\, a=\const~. 
\lb{sigma_cl}  
\ee   
As the constant $a$ in eq.\rf{sigma_cl} varies from minus infinity to $L/2$\,,
$\sigma(L)$ stays exponentially, {\it i.e.\/}\ up to $O(\e{-L})$, close to one,
whereas $\sigma(0)$ goes from $1$ to $-1\,{+}\,O(\e{-L})$, as illustrated by Fig.~1. 
Therefore, up to unessential exponentially small corrections, we can satisfy the 
boundary conditions\rf{bc} by the solutions\rf{sigma_cl}
where the parameter $a$ is taken to be the following function of $s$: 
\be
a=\ln\left(\frac{1-s}{1+s}\right)~.
\lb{as}
\ee

When the variation in $s$ drives the system from the trivial sector to the topological one,
the lowest of the normal frequencies\rf{vac_mod} decreases from 
$\omega^{(0)}_0\rav \sqrt{1+\left(\frac{\pi}{L}\right)^2}\,{\simeq}\,1$
to an infinitesimal value for $L\sh{\to}\infty$.  The
zero point oscillations in the lowest normal mode become more and more 
broad until the quantum ground state of the system completely de-localizes.
By this moment, the quasi-classical description of the corresponding normal coordinate
breaks down, and it should be replaced by the collective quantization as $s\sh{\to}{-}1$.
The other oscillation modes retain
finite positive frequencies and their quasi-classical description holds all the time.
Thus the first-order quantum correction to the soliton mass, $M$,
may be determined as the sum of the shifts $\frac12\,\Delta\omega'_k\rav
\frac12\,\omega'_k\sh{-}\frac12\,\omega^{(0)}_k$ for the $k\sh{\ge}1$ modes
{\it minus\/} half of the frequency of the disappeared lowest vacuum mode.
This is evident from the 
requirement of having an equal number of degrees of freedom, whether collectively 
quantized or not, in both sectors of the regularized theory. 
For notational convenience, let us define $\Delta\omega'_0\sh{\equiv}\omega'_0\sh{-}
\omega^{(0)}_0$ where $\omega'_0\sh{\equiv}0$ corresponds to the zero eigenvalue
of eq.\rf{nonloc2}, so that 
\be
M_{\rm 1-loop}=\frac12\sum_{n=0}^{\infty}\Delta\omega'_k +\mbox{c.t.}~.
\nn\ee 
In this equation ``c.t.'' stands for the counter terms contribution due to 
one-loop renormalization of parameters in the theory.  We will return to it shortly.

By now, regularization of $E'_{\rm Cas}[\sigma]$ should become apparent. 
Indeed, from the physical interpretation of the reduced effective energy,
eq.\rf{min_sigma}, we conclude that at its minimum,~$\sigma_0$, 
\be
E'_{\rm Cas}[\sigma_0] = M_{\rm 1-loop} =
\frac12\sum_{n=0}^{\infty}\Delta\omega'_k[\sigma_0] +\mbox{c.t.}[\sigma_0]~.
\lb{ren_fix}
\ee
Since all the vacuum subtractions needed to regulate $E'_{\rm Cas}[\sigma]$,
together with the unphysical $\omega'_0[\sigma]\sh{\equiv}0$, do not depend 
on the function~$\sigma(x)$ and we determined them for some particular 
configuration~$\sigma_0(x)$, they simply carry over to an arbitrary $\sigma(x)$ in the 
same topological sector as  
\be
E'_{\rm Cas}[\sigma] = 
\frac12\sum_{n=0}^{\infty}\Delta\omega'_k[\sigma] +\mbox{c.t.}[\sigma]~,\lb{ECas1}\\
\Delta\omega'_k[\sigma] \equiv \omega'_k[\sigma] - \omega^{(0)}_k~. \qquad 
\lb{d_omega}
\ee 

Let us assume for simplicity that the Lagrangian\rf{Lagr} is invariant under the 
$\phi\sh{\to}{-}\phi$ transformation and take $\sigma_1\rav{-}\sigma_2$ for the
$\sigma(x)$ boundary conditions\rf{sigmabc}.
Then it is sufficient to consider only the configurations
$\sigma(x)$ that are odd functions of $x$ so that the eigen-functions  $\eta_n(x)$ in
eq.~(\ref{nonloc2}) are either even or odd. 
Of course, now we imply that our regularization methods
also respect parity, for example, the coordinate $x$ varies from $-L$ to $L$,
where $L\sh{\gg}\frac1m$\,.
As $x\sh{\gg}\frac1m$\,, $\eta_n(x)$ approaches the asymptotic form $\cos(k_n x+\delta'_+)$
or $\sin(k_n x+\delta'_-)$ for even or odd functions respectively with
$k_n\rav\sqrt{\omega'_n{}^2\sh{-}m^2}$\,.  The frequency shifts\rf{d_omega} are
related to the scattering phases $\delta'_+(k_n)$ or $\delta'_-(k_n)$ as
\be
\Delta\omega'=\frac{d\omega'}{dk}\,\Delta k = 
        \frac{d\omega'}{dk}\left(-\frac{\delta'_{\pm}(k)}{L}\right)~,
\nn\ee
because $kL\rav(k\sh{+}\Delta k)L+\delta'_{\pm}(k)$,  provided $k\sh{\gg}1/L$. 
This part of the spectrum of eq.\rf{nonloc2} contributes to the sum\rf{ECas1} as
\be
E_{\rm Cas}^{'(cont)}&=&\frac12\int\left(\frac{dkL}{\pi}\right)
     \left(\frac{d\omega'}{dk}\right)\left(-\frac{\delta'_+(k) +\delta'_-(k)}{L}\right)=
\nonumber\\
     &=& -\,\frac{1}{2\pi}\int\frac{k\,dk}{\sqrt{k^2+m^2}}\left[\delta'_++\delta'_-\right]~.
\lb{ECasc}
\ee
The spectrum of eq.~(\ref{nonloc2}) will also have a few discrete modes with 
$\omega'_n\sh{<}m$ and the corresponding $\omega^{(0)}_n\rav m\sh{+}O(L^{-2})$.
Summing up all the contributions,
\be
E'_{\rm Cas}=\frac12\sum_{\rm discrete\atop \rm spectrum}(\omega'_n-m)-
 \frac{1}{2\pi}\int^{\infty}_0\frac{k\,dk}{\sqrt{k^2+m^2}}\left(\delta'_+ +\delta'_-\right)
 +\mbox{c.t.}~.
\lb{ECas2}
\ee 
The integral in eq.\rf{ECasc} is a poor approximation when $k\,{\le}\,L^{-1}$
but because for such modes\footnote{Unless the state becomes
bound when it is counted separately.
}
$\Delta\omega'\rav O(L^{-2})$ and their number is $O(L^0)$, we
can safely set the lower limit of $dk$ integration to zero, provided the
integral converges in the infra-red.

Since the scattering phases $\delta'_{\pm}$ fall off as $1/k$ for large momenta,
the integral in eq.\rf{ECas2} still diverges logarithmically on the upper limit. 
This ultra-violet divergence is removed by the usual parameter renormalization that 
manifests itself in the form of the counter term in eq.\rf{ECas2}. Let us remember that 
counterterms are conventionally defined by a certain renormalization condition in the 
perturbative sector of the theory independently of the collective quantization
procedure. 
By translational invariance, the corresponding term in the action, for instance,
\be
S_{\rm c.t.}^{(\phi^4)}\propto \int d^2 x\left[\phi^2(x)-1\right]
\lb{Sct}
\ee
in the $\phi^4$ theory, can not depend on the collective coordinate~$X$.
The total momentum $P$ is also $g$-suppressed in $S_{\rm c.t.}[\phi]$, as
in eq.\rf{Hq}, or does not appear at all in the absence of wave function renormalization
that is the case in our example\rf{Sct}. 
Therefore, $S_{\rm c.t.}\rav S_{\rm c.t.}[q]$ simply modifies the potential $V[q]$ 
in eq.\rf{Hq}, and for the purpose of obtaining the explicit form of the counterterm 
contribution to the reduced effective energy,  we may ignore the reduction to the $P\rav 0$ 
sector and consider the conventional effective energy $E_{\rm eff}\rav\Gamma/T$ where 
$\Gamma[\phi]$ is the usual effective action 
describing a spatially localized field configuration $\phi(x)\rav g^{-1}\sigma(x)$. 

The one-loop part of $\Gamma$ in the theory\rf{Lagr} is of the form
\be
\Gamma_{\rm 1-loop}[\phi]&=& i\ln\det\left(\frac{\df^2 S}{\df \phi^2}\right)=
  i\tr\ln\left(-\d^2-\U''\left(g\phi\right)\right)=\\
  &=&i\tr\ln\left(-\d^2-m^2-v\right)
~~,~~\,
v(g\phi(x))\equiv \U''\left(g\phi(x)\right)-m^2~,  \lb{Dv}
\nn\ee
where again $m^2$ stands for the value of $\U''$ at its minimum 
and describes the mass of the perturbative plane wave excitations over the true vacuum.
Denoting the operator $(-\d^2\,{-}\,m^2)$ by $\Delta_0$,
we have the following expansion in powers of the ``interaction''~$v(\phi)$:
\be
\Gamma_{\rm 1-loop}[\phi]= i\tr\ln\Delta_0 - i\tr\left(\Delta_0^{-1}v\right)
      - i\tr\left(\Delta_0^{-1}v \Delta_0^{-1}v\right) - \dots~.
\lb{loop_expansion}
\ee
The first term in this expansion is a constant independent of the field 
configuration $\phi$ that is trivially removed by setting the vacuum energy to zero. 
The rest of the terms may be presented graphically by the sum
of the diagrams shown on Fig.~2.
\begin{figure}[t]
\centering
\leavevmode
\epsfbox{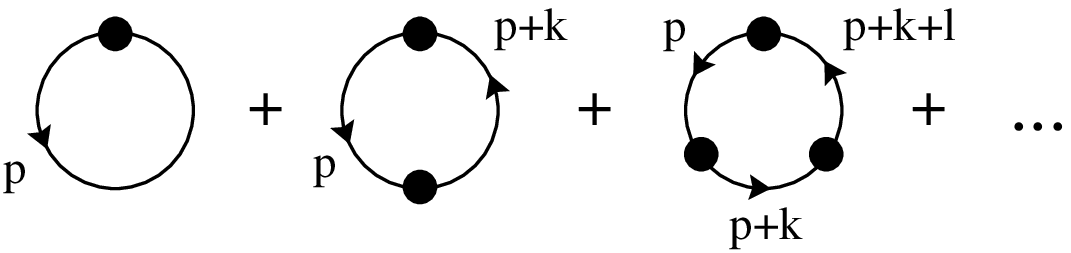}
\podpis{The diagrammatic presentation of the one-loop contribution
         to the effective action,~$\Gamma_{\rm 1-loop}$. In the momentum representation,
         a line carrying a momentum $p$ stands for the usual free boson 
         propagator $\frac{1}{p^2-m^2}$, and the insertions are given by  
         $\tilde v(k)\rav\int d^2x\,\e{ikx}v(x)$ where $k$ is the injected
         momentum. The integration goes over all the loop and the injected
         momenta, {\it e.g.\/} over $p$, $k$, and~$l$ for the third graph above.
} 
\label{fig2}
\end{figure}
The inclusion of $\Gamma_{\rm c.t.}[\phi]\rav {-}S_{\rm c.t.}[\phi]$ should remove all
the divergences from the loop integrals on Fig.~2. 
For a scalar field in one dimension this can be achieved$^{[7]}$
by modifying the 
scattering phases in eq.\rf{ECas2} as 
\be
\delta'_{\pm} \rightarrow ~\delta'_{\pm} - \delta^{(1)}_{\pm}~. 
\nn\ee
Here $\delta^{(1)}_{+}$ and $\delta^{(1)}_{-}$ are the first Born approximations to the 
scattering phases
in the continuous spectrum of the oscillation modes about a ``nailed down'' 
soliton problem\footnote{
This trick works because the consecutive Born approximations for eq.\rf{loc} and the 
expansion\rf{loop_expansion} are formulated as a power series of same quantity,~$v$\,. 
Similar methods can be carried out for sufficiently general theories 
of bosonic or fermionic fields in the realistic three space dimensions$^{[16]}$.
},
\be
\omega^2\eta(x)=
   \left[-\,\frac{d^2}{dx^2}+\U''(\sigma)\right]{\eta}(x)=
   \left[-\,\frac{d^2}{dx^2}+m^2+v(\sigma(x))\right]{\eta}(x)~.
\lb{loc}
\ee
This operation is equivalent to adding a counter term that cancels the shift of
the field vacuum expectation value in the perturbative sector due to one-loop corrections.
The first Born approximations to the scattering phases in eq.\rf{loc} are
easily found$^{[7]}$
as
\be
\delta^{(1)}_+(k)&=&-\,\frac{1}{k}\int_{0}^{+\infty}dx\,v(x)\cos^2{kx}~\nn \\ 
\delta^{(1)}_-(k)&=&-\,\frac{1}{k}\int_{0}^{+\infty}dx\,v(x)\sin^2{kx}~,
\nn\ee
and our final formula for the Casimir energy of the collectively quantized
system reads~:
\be
E'_{\rm Cas}&=&\frac12\sum_{\rm discrete\atop \rm spectrum}(\omega'_n-m)
  -\frac{1}{2\pi}\int_0^{+\infty}\frac{k\,dk}{\sqrt{k^2+m^2}}~\delta'_{tot}(k)~,
									\lb{ECasf}\\
\mbox{where}&&
\delta'_{tot}(k)\equiv \delta'_+(k)+\delta'_-(k)+
\frac{1}{k}\int_{0}^{+\infty}dx\,v(x)~.
\lb{delta_t}
\ee
It should be reassuring that the non-local separable potential in eq.\rf{nonloc2}
does not affect the $1/k$ asymptotic behavior of $\delta'_{\pm}(k)$ and the 
sum\rf{delta_t} falls off as $1/k^2$ yielding a finite integral in eq.\rf{ECasf}.

\section{Numerical Computations}

As a demonstration of our method, we calculate the reduced effective energy for
a set of field configurations in the one-dimensional $\phi^4$ example
\be
\L=\frac{1}{2}\,\left({\d_{\mu} \phi}\right)^2
{-}~\frac{1}{8g^{2}}\left((g\phi)^2-1\right)^2+\L_{ct}~\,.
\lb{Lphi4}
\ee

We exploit two different families of trial configurations.  The first one is the scaling
transformation of the classical kink solution\rf{sigma_cl}:
\be
\sigma_{\rm I}(x)\equiv g\phi(x) = \tanh\frac{x}{2\w}~. 
\lb{trial1}
\ee
The parameter $\w$ characterizes the width of the kink\rf{trial1} 
relative to its classical value.  As it is seen from further calculations,
the scaling transformation\rf{trial1} possesses some unexpected peculiar properties.
For this reason we also consider an alternative variation of the solution\rf{sigma_cl}
constructed to represent some ``absolutely non-special'' direction in the functional
space of all admissible variations:
\be
\sigma_{\rm II}(x) = \tanh\frac{x}{2} + 
             \alpha x\exp{\left(-\frac{\pi x^2}{2}\right)}~.
\lb{trial2}
\ee

The classical contribution to the effective energy, $E_{\rm cl}[\sigma]$, is trivially
calculated from the formula\rf{Eclass}. Of course, the result should take the minimum, 
$M_{\rm cl}\rav\frac{2}{3g^2}$, at the classical values of the variational parameters 
$\w\rav 1$ or $\alpha\rav 0$ and grow quadratically with $\delta\omega$ and 
$\delta\alpha$, for example, 
\be
E_{\rm cl}[\sigma_{\rm I}(x)]=\frac{1}{3g^2}\left(\w+\frac{1}{\w}\right)=
  \frac{2+\delta\w^2+O(\delta\w^3)}{3g^2}
  ~~,\quad\delta\w\equiv \w-1~.
\lb{EclI}
\ee  

According to the previous sections, the reduced Casimir energy is determined
by eqs.~(\ref{ECasf}-\ref{delta_t}) where the discrete oscillation mode 
frequencies,~$\omega'_n$, and the scattering phases in the continuous spectrum, 
$\delta'_{\pm}(k)$, refer to the solutions of the auxiliary eq.~(\ref{nonloc2}).
The following Subsection~6.1 describes how to actually solve these non-local
equations and extract the required quantities $\omega'_n$ and~$\delta'_{\pm}(k)$.
The reader who is not interested in these details should skip to Subsection~6.2
where we analyze our numerical results.

\subsection{Practical Implementation}

Substituting the explicit form of the potential~$\U$, eq.\rf{Uphi4}, let us rewrite 
eq.~(\ref{nonloc2}) as
\be
-\eta''(x)+v(x)\eta(x)+a_i(x)\int_{-\infty}^{+\infty} dy~b_i(y)\,\eta(y)=
       \left(\omega'^2-1\right)\eta(x)~,  
\lb{main_eq}
\ee
where we sum over the repeated index $i\rav 1,2$ and define
\be
v(x)\equiv \U''(\sigma(x))-1 = \frac{3}{2}\left(\sigma^2(x)-1\right)~,
\nn\ee
\be
a_1(x)\equiv \frac{1}{\sqrt{I_{\sigma}}}\,j_{\sigma}'(x)&,&
b_1(y)\equiv \frac{1}{\sqrt{I_{\sigma}}}\,\sigma'(y)~,\\
a_2(x)\equiv \frac{1}{\sqrt{I_{\sigma}}}\,\sigma'(x)&,&
b_2(y)\equiv \left[\frac{1}{\sqrt{I_{\sigma}}}\,j_{\sigma}'(y)-
\frac{I_J}{I_{\sigma}\sqrt{I_{\sigma}}}\, \sigma'(y)\right]
\nn\ee
with $
I_{\sigma}\equiv \int dz\left(\sigma'\right)^2,
       ~I_J\equiv \int dz\left(j_{\sigma}'\sigma'\right)\,.
$

If the configuration $\sigma(x)$ is an odd function of $x$, which is true in our 
cases~(\ref{trial1},\,\ref{trial2}), the functions $v(x)$, $a_i(x)$, and $b_i(x)$ 
are even and the modes $\{\eta(x)\}$ divide into the parity even (symmetric) and odd 
(antisymmetric) channels. The separable potential term in eq.\rf{main_eq} does not 
appear in the antisymmetric channel because the integral over $dy$ trivially vanishes 
for the odd $\eta(y)$. Changing the lower integration limit to zero in 
the symmetric channel, we reformulate the problem\rf{main_eq} as:
\be
\left\{\begin{array}{l}
K_{\omega'}\eta_+(x)+2a_i(x)\int_{0}^{+\infty} dy~b_i(y)\,\eta_+(y)=0~\\
\mbox{with}\quad\eta'_+(0)=0~
\end{array}\right.
&&\quad \mbox{(symmetric channel)}~, \lb{sp}\\
\left\{\begin{array}{l}
K_{\omega'}\eta_-(x)=0~\\
\mbox{with}\quad\eta_-(0)=0~ 
\end{array}\right.
\qquad\qquad&& \mbox{(antisymmetric channel)} 
\lb{ap}
\ee
where $K_{\omega'}$ is the local linear differential operator
\be
K_{\omega'} = -\,\frac{d^2}{dx^2}+v(x)-(\omega'^2-1)~.
\nn\ee

First, let us determine the discrete spectrum,
formed by the solutions with $\omega'\sh{<}1$ that exponentially vanish as 
$x\sh{\to}{+}\infty$\,.  In the antisymmetric channel we simply integrate the 
differential equation\rf{ap} from $0$ to $+\infty$ with initial conditions
$\eta_-(0)\rav0$, $\eta'_-(0)\rav1$ and search for $\omega'$s for which $\eta_-(x)$ 
falls off exponentially at large~$x$\,.  Of course, this has to be repeated for every 
configuration~$\sigma(x)$. 

In the symmetric channel, eq.\rf{sp}, we can
also apply this shooting technique looking for the solutions in the
form
\be
\eta_+(x)=\eta_0(x) + \alpha_i \eta_i(x)~,
\lb{eta_ansatz}  
\ee
where we sum over $i\rav 1,2$\,. In this ansatz
$\eta_0$, $\eta_1$, and $\eta_2$ solve the ordinary differential 
equations 
\be
K_{\omega'}\eta_q(x)\equiv -\eta''_q(x)+v(x)\eta_q(x)-(\omega'^2-1)\eta_q(x)=
\left\{
\begin{array}{cc}
0 & \mbox{for}~~q=0\\
a_q(x) & \mbox{for}~~q\equiv i= 1,2
\end{array}\right.
\nn\ee
with the boundary conditions
\be
\eta_0(0)=1\,,~\eta'_0(0)=0~;\quad \eta_i(0)\,,\,\eta'_i(0)=0~.
\nn\ee
Substituting the ansatz\rf{eta_ansatz} into eq.\rf{sp} we see that the latter holds
provided the coefficients $\alpha_1$ and $\alpha_2$ are determined from
the following linear algebraic equation:
\be
\alpha_i+\beta_i+\gamma_{ij}\,\alpha_j = 0~,
\lb{alpha}
\ee
where again $i$ and~$j$ range over 1 and~2, and
\be
\beta_i\equiv 2\int_{0}^{+\infty} dy~b_i(y)\,\eta_0(y)~,~~
\gamma_{ij}\equiv 2\int_{0}^{+\infty} dy~b_i(y)\,\eta_j(y)~.
\nn\ee
The outcome of the described procedure applied to our first family of the trial 
configurations, eq.\rf{trial1},  in both symmetric and antisymmetric
channels is presented on Fig.~3\,a) (solid lines). 
We would like to postpone its discussion until the following subsection.

In the continuous spectrum of the symmetric channel we look for the solution 
of eq.\rf{sp} in the form 
\be
\eta_+(x)=\e{i\delta'_+(k)}\varphi(x)+\e{-i\delta'_+(k)}\varphi^*(x)
\nn\ee
where $\varphi(x)$ is a formal solution of eq.\rf{sp} for $x\in [0,+\infty)$ such that 
\be
\varphi(x) \rightarrow \e{ikx}\quad\mbox{as}\quad x\to+\infty~,
\nn\ee
and the scattering phase $\delta'_+(k)$ is adjusted to satisfy the condition
\be
\eta_+'(0)=0~.
\lb{sbc}
\ee
In order to achieve better numerical precision, we trade the oscillating function
$\varphi(x)$ for a smoother one, $c(x)$, as 
\be
\varphi(x)=c(x)\e{ikx}~
\qquad \mbox{with} \qquad c(+\infty)=1~~\mbox{and}~~c'(+\infty)=0~
\nn\ee
(we remember that $k\sh{\equiv}\sqrt{\omega'^2-1}$ and $c(x)$ depends on
the parameter $k$ as well).
The function~$c(x)$ also satisfies a linear differential equation with a separable 
potential,
\be
-\,c''(x)-2ikc'(x)+v(x)c(x)+2\,\e{-ikx}a_i(x)\int_0^{+\infty} dy~b_i(y)\e{iky}c(y)=0~,
\lb{sepeq}
\nn\ee
and is determined, just as in the case of eq.\rf{sp}, with the ansatz
\be
c(x)=c_0(x) + \tilde\alpha_i c_i(x)~,
\lb{c_ansatz}  
\ee
where
\be
-\,c''_q(x)-2ikc'_q(x)+v(x)c_q(x) = \left\{
\begin{array}{cc}
0~ & \mbox{for}~~q=0~,~\\
a_q(x)\,\e{-ikx} & \mbox{for}~~q=1,2~;
\end{array}\right.
\nn\ee
\be
c_0(+\infty)=1\,,~c'_0(+\infty)=0~;\quad c_i(+\infty)\,,\,c'_i(+\infty)=0~.
\nn\ee
Again, we have the algebraic equation $\tilde\alpha_i\sh{+}\tilde\beta_i\sh{+}
\tilde\gamma_{ij}\,\tilde\alpha_j
\rav 0$ for the coefficients $\tilde\alpha_1$ and $\tilde\alpha_2$ where
\be
\tilde\beta_i= 2\int_0^{+\infty} dy~b_i(y)\e{iky}\,c_0(y)~,~~
\tilde\gamma_{ij}= 2\int_0^{+\infty} dy~b_i(y)\e{iky}\,c_j(y)~.
\nn\ee
Having calculated the function $c(x)$, we find the phase~$\delta'_+(k)$ from the
condition\rf{sbc} as
\be
\delta'_+=-\arg\,[k\,c(0)-ic'(0)]~,
\nn\ee
where the branch of the argument functions is unambiguously specified by the 
requirement $\lim_{k\to{\infty}}\delta'_+\rav 0$~.

The antisymmetric channel is trivial:  The continuous solutions of eq.\rf{ap}
are given by
\be
\eta_-(x)=\e{i\delta'_-(k)}{\varphi}(x)-\e{-i\delta'_-(k)}
               \varphi^*(x)&&\left(~\mbox{such that}~\eta_-^{}(0)=0~\right)~,
\nn\ee
where
\be
\varphi(x)=c(x)\e{ikx}~\qquad \mbox{with} \qquad c(+\infty)=1\,,~c'(+\infty)=0
\nn\ee
and
\be
-~c''(x)-2ikc'(x)+v(x)c(x)=0~.
\nn\ee
From the requirement $\eta_{-}(0)\rav 0$ we obtain
\be
\delta'_-=-\arg\,c(0)~
\nn\ee
in the argument branch such that $\lim_{k\to{\infty}}\delta'_- \rav 0$\,.
Calculating the scattering phases in both channels and regularizing them by removing 
the first Born approximation, eq.\rf{delta_t}, we obtain the continuous spectrum 
contribution to the Casimir energy.

\subsection{Discussion of the Results}

\begin{figure}[t]
\centering
\leavevmode
\epsfbox{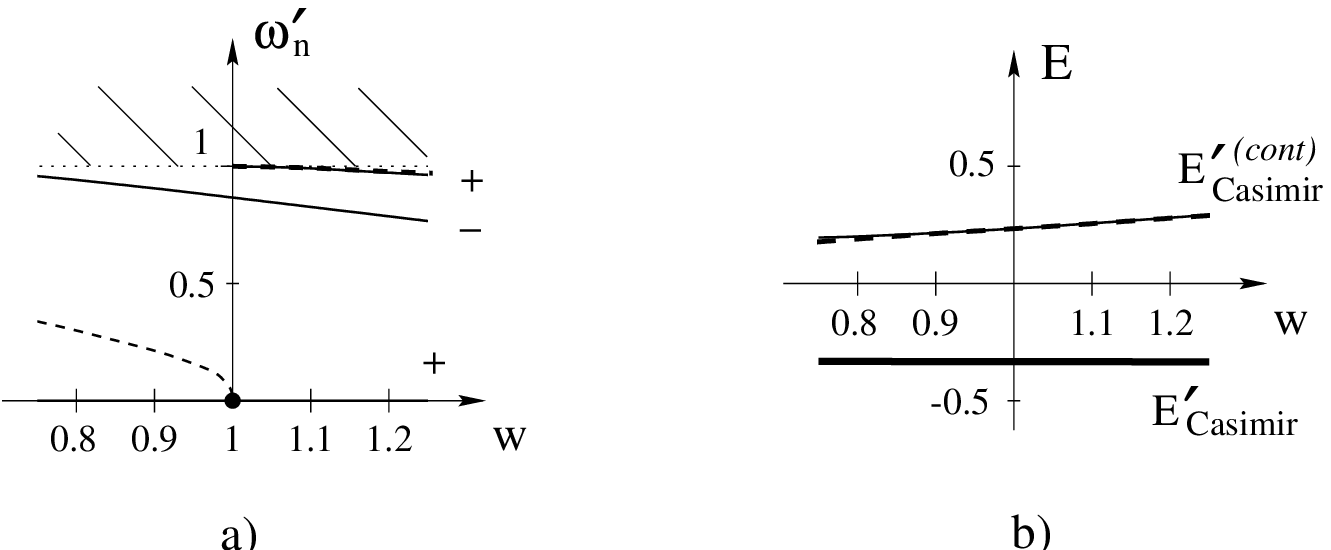}
\podpis{ The normal oscillation frequencies in the discrete 
spectrum,~(a), and the reduced Casimir energy together with the contribution to it from the 
continuous spectrum,~(b), versus the soliton width, $\w$, in the ansatz $\sigma_I\rav 
\tanh\frac{x}{2\w}$\,. The solid lines on the plots describe $\omega'_n$,
$E'^{(cont)}_{\rm Cas}$, and $E'_{\rm Cas}$ in the theory reduced to the $P\rav 0$
sector. The dashed lines are the results for $\omega_n$ and $E^{(cont)}_{\rm Cas}$   
in a conventional analysis of small oscillations about a kink with a definite
position in space. The only drastic difference between the two approaches occur in the
translational mode, for which $\omega'_0\sh{\equiv}0$, but $\omega_0\sh{\propto}
\sqrt{1-\w}$ and is imaginary when $\w\sh{>}1$\,.
}
\end{figure}

The spectrum of small oscillations about the classical kink solution 
$\sigma_{\rm cl}\rav\tanh{\frac{x}{2}}$ consists of the continuous spectrum
with $\omega\rav\sqrt{1+k^2}$, a parity odd discrete state at 
$\omega_-\rav\frac{\sqrt3}{2}$, and the even $\omega_0\rav0$ translational mode.
Besides, the parity even amplitude $\eta_+(x)$ corresponding to $\omega_+\rav1$ approaches
a constant value as $x\sh{\to}\infty$. Infinitesimally lowering
the frequency with an arbitrarily small variation of the
soliton shape, one would send this state into the discrete spectrum.
It is often convenient to refer such a state as a half-bound one. 

A small variation of the shape function induces a shift in the 
oscillation frequencies and the total Casimir energy. In general, the shifts are 
expected to be linear in the variational parameters, $\delta\w\rav\w\sh{-}1$ or 
$\alpha$ in our examples\rf{trial1} or\rf{trial2}, of course, up to 
less important in the perturbative regime higher power corrections. The
non-vanishing of $\left.\frac{d\omega'}{d(\delta\w)}\right|_{\delta\w\rav0}$ is clearly 
visible for the odd discrete state on Fig.~3\,a) and for the continuous contribution 
to the Casimir energy on Fig.~3\,b), where the ansatz\rf{trial1} was considered. 
The dashed lines on this Figure present the same quantities computed for
a localized kink configuration$^{[7,10]}$,
when the non-local terms in eq.\rf{nonloc2} 
are absent. The discrepancy between the two methods for the non-zero modes
is barely observable, 
demonstrating the weakness of the collective quantization influence on these
modes in the perturbative regime.

As expected, there always exists a normalizable zero-frequency solution in the
symmetric channel of the reduced problem\rf{nonloc2}.  For a localized configuration, 
the ``zero''-mode frequency  would vary as $\omega_0\sh{\propto}\sqrt{1\sh{-}\w}$ 
and become imaginary as $\w$ exceeds~$1$\,.
This square root singularity and the imaginary contribution to the conventional
effective energy has nothing to do with the soliton instability (it is stable!).
It is rather due to inadequacy of the standard effective energy formalism in 
describing the true solitonic ground state that has no specified position in space.
 
The wide solid line on Fig.~3\,b) is our final result for $E'_{\rm Cas}$ verses the kink 
width.  The first perturbative correction to the soliton mass$^{[1,17,10]}$,
\be 
\Delta M_{1-loop}=E^{\rm (\prime)}_{\rm Cas}(0)= \frac1{4\sqrt{3}}-\frac3{2\pi} \simeq
-0.333127~, 
\lb{dM_th}
\ee
is numerically found to be $-0.333124$, which, given our numerical accuracy, is 
in complete agreement with the analytical result\rf{dM_th}.  
Surprisingly, the plot on Fig.~3\,b) shows no apparent dependence of $E'_{\rm Cas}$
on the width~$\w$. 
\begin{table}[t]
\begin{tabular}{|c|llllll|}
\hline
$\w-1$  &~~$-$0.1 &~\,$-$0.05 &~$-$0.025 &~~~0~ &~\fmin 0.025 &~\,\fmin 0.05 \\ 
\hline
$\frac12\,\omega^{(\prime)}_- - \frac{\sqrt3}{4}$ 
&\fmin 0.0202 &\fmin 0.0101 &\fmin 0.0050 &~~~0~ &$-$0.0050 &$-$0.0100 \\
$E^{(cont)}_{\rm Cas} - (1-\frac1{2\sqrt3}-\frac3{2\pi})$ 
&$-$0.0211 &$-$0.0102 &$-$0.0050 &~~~0~ &\fmin 0.0051 &\fmin 0.0104 \\
$E'^{(cont)}_{\rm Cas} - E^{(cont)}_{\rm Cas}$ 
&\fmin 0.0023 &\fmin 0.0005 &\fmin 0.0001 &~~~0~ &\fmin 0.0001 &\fmin 0.0003 \\
\hline
$E'_{\rm Cas} - (\frac1{4\sqrt3}-\frac3{2\pi})$ 
&\fmin 0.0014 &\fmin 0.0004 &\fmin 0.0001 &~~~0~ &\fmin 0.0002$^*$ &\fmin 0.0006$^*$ \\
\hline
\end{tabular}
\podpis{ The change in the reduced Casimir energy (the last line) against the soliton 
width variation and the main three contributions to it.  The results
demonstrate that, up to insignificant higher power corrections, 
$\delta\omega^{(\prime)}_-\sh{\propto}(\w\sh{-}1)$, 
$E^{(cont)}_{\rm Cas}\sh{\propto}(\w\sh{-}1)$, and 
$(E'^{(cont)}_{\rm Cas}\sh{-} E^{(cont)}_{\rm Cas})\sh{\propto}(\w\sh{-}1)^2$\,. 
The last two values for $E'_{\rm Cas}$, marked by an asterisk, should also include 
the symmetric discrete mode with the frequency very close to one.
}
\end{table}
Indeed, in Table~1 we give the deviations of numerically computed contributions to
$E'_{\rm Cas}$ from their classical values for the antisymmetric discrete mode 
$\omega'_-(\w)$, which coincides with $\omega_-(\w)$, and for the sum of the continuous 
modes, $E'^{(cont)}_{\rm Cas}$, the latter being separated into $E^{(cont)}_{\rm Cas}$
obtained for a localized soliton and the correction to it due to collective 
quantization.  As one can see from the table, the linear variation in the discrete
spectrum exactly cancels by the local part of the continuous spectrum and one is
left with the quadratic in $\delta\w$ dependence, which is largely produced by the
non-local terms in eq.\rf{nonloc2}. Thus our calculations find {\it no} one-loop
quantum correction to the soliton width, $\w$.  This result comes unexpected, 
remembering that $\sigma(x)$ should receive non-vanishing one-loop correction
according to eq.\rf{d_sigma}. We do find such a correction when we consider another 
family of trial configurations, $\{\sigma_{II}(x)\}$ in eq.\rf{trial2}. 
\begin{table}[t]
\begin{tabular}{|c|llllll|}
\hline
$\alpha$ &~\,$-$0.05 &~$-$0.025 &~~~0~~ &~\fmin 0.025 &~\,\fmin 0.05 &~~\fmin 0.1 \\ 
\hline
$\frac12\,\omega^{(\prime)}_- - \frac{\sqrt3}{4}$
&$-$0.00091 &$-$0.00046&~~~0~~ &\fmin 0.00046 &\fmin 0.00092 &\fmin 0.00185 \\
$E^{(cont)}_{\rm Cas} - (1-\frac1{2\sqrt3}-\frac3{2\pi})$ 
&\fmin 0.0018 &\fmin 0.0009 &~~~0~~ &$-$0.0009 &$-$0.0018 &$-$0.0038 \\
$E'^{(cont)}_{\rm Cas} - E^{(cont)}_{\rm Cas}$ 
&\fmin 0.0053 &\fmin 0.0013 &~~~0~~ &\fmin 0.0013 &\fmin 0.0053 &\fmin 0.0201 \\
\hline
$E'_{\rm Cas} - (\frac1{4\sqrt3}-\frac3{2\pi})$ 
&\fmin 0.0062$^*$ &\fmin 0.0017$^*$ &~~~0~~ &\fmin 0.0009 &\fmin 0.0044 
&\fmin 0.0181 \\
\hline
\end{tabular}
\podpis{ The numerical results for the second family of trial 
configurations, $\{\sigma_{II}(x)\}$.}
\end{table}
The numerical results
for this choice are summarized in Table~2. We obtain that 
\be
\left.\frac{\partial E'_{\rm Cas}}{\partial \alpha}\,\right|_{\alpha=0}=
\,\left.\frac{\partial E_{\rm Cas}}{\partial \alpha}\,\right|_{\alpha=0}\simeq 
(\,-1.5\pm 0.1\,)\times10^{-2}~,
\nn\ee 
resulting in 
\be
\delta\alpha_{1-loop}= 
      \left.-\,\frac{\left(\frac{\partial E'_{\rm Cas}}{\partial \alpha}\right)}
     {\left(\frac{\partial ^2E_{\rm cl}}{\partial \alpha^2}\right)}\,\right|_{\alpha=0}=
     g^2\cdot(\,2.2\pm 0.2\,)\times10^{-2}~.
\nn\ee
We see that the first quantum correction to the shape parameter $\alpha$ is small until
$g^2\sh{\ll} 10^2$\,. However, the contribution of the non-local terms becomes 
significant at a much smaller coupling. From our calculations,
\be
E'^{(cont)}_{\rm Cas} - E^{(cont)}_{\rm Cas} \simeq \frac{(\,4.2\pm 0.1\,)\,\alpha^2}{2}~,
\lb{Casstab}
\ee 
whereas $\frac{\partial ^2E_{\rm cl}}{\partial \alpha^2}\sh{\simeq} 0.6956/g^2$~.
The effect of the quantum correction\rf{Casstab} becomes comparable with the 
classical energy at $g^2\sh{\sim}\frac{0.6956}{4.2}\sh{\sim}0.17$\,.

\section{Conclusion}

To summarize, we adapted the conventional effective energy variational techniques
to the situation where the Hamiltonian of the system possesses cyclic variables 
and the standard semi-classical methods should incorporate their collective quantization.
In our formalism one may retain the description of the system in terms of a continuous
field,~$\phi(x)$, by introducing an external source with a special non-linear
non-local coupling. The source term in eq.\rf{Haux} insures the coupling to all the normal 
degrees of freedom excluding the cyclic variables. 
The effective energy functional reduced to 
the $P\rav0$ sector of solitonic states, $E'_{\rm eff}[\sigma(x)]$, is given to 
first order in $\hbar$ by the classical energy of the configuration 
$\phi(x)\rav g^{-1}\sigma(x)$ and the properly regularized reduced Casimir 
energy,~eq.\rf{ECasf}. The mass of the soliton in the ground state equals the
minimum of $E'_{\rm eff}[\sigma]$, and the configuration $\sigma(x)$ at which
this minimum is achieved describes the soliton form-factor, eq.\rf{formfr}.

The effect of collective quantization technically amounts to the introduction 
of the non-local separable potential\footnote{
Some aspects of the Casimir energy for systems described
by separable potentials has been considered by Jaffe and Williamson$^{[18]}$.
}   
in the eigen-mode equation\rf{nonloc2}. 
However, in order to obtain the first quantum 
corrections to the {\it classical} values of the soliton mass and form-factor, one could
ignore these non-local terms but follow the rule that the lowest oscillation 
eigen-frequency must be excluded from the Casimir energy.
The correction to the reduced energy due to collective quantization grows 
quadratically with the deviation from the classical field configuration.
To get a possible insight into the physics in the non-perturbative regime,
one may still consider $E'_{\rm eff}[\sigma(x)]$ computed only to the one-loop
order. This can be done numerically for an arbitrary field configuration
as described in Section~5.

It would be interesting to apply this method to investigation of the role of 
collective quantization in deeply non-perturbative regime.  This could be 
necessary, for example, for a consistent treatment of the
models$^{[19]}$
in which a non-classical field configuration 
is stabilized by an order $\hbar$ contribution to the soliton energy from fermionic 
fields coupled to the scalar ``Higgs''.  An extension of the method that would 
incorporate the fermionic fields as well, presents another challenge by itself.  

\section{Acknowledgments}

I would like to thank 
E.~Farhi, J.~Goldstone, N.~Graham, and H.~Weigel for numerous fruitful discussions.
My special thanks to Robert~L.~Jaffe for pointing my attention to this problem and
for his continuous interest and helpful suggestions.


\begin{thebibliography}{99}
\bibitem{Cas_proof}
The proof for the usual effective energy is given, for example, in Section~5 of~\cite{R}
and is straightforwardly generalizable to our situation.
\end{thebibliography}
\end{document}